\documentclass[conference]{IEEEtran}
\usepackage{dannychen}
\IEEEoverridecommandlockouts

\date{} %leave blank

\begin{document}
\title{Efficient Quantum Circuit Cutting by Neglecting Basis Elements}

\author{\IEEEauthorblockN{Daniel T. Chen\IEEEauthorrefmark{1}, Ethan H. Hansen\IEEEauthorrefmark{1}, Xinpeng Li\IEEEauthorrefmark{1}, Vinooth Kulkarni\IEEEauthorrefmark{1}, \\ Vipin Chaudhary\IEEEauthorrefmark{1}, Bin Ren\IEEEauthorrefmark{3}, Qiang Guan\IEEEauthorrefmark{2}, Sanmukh Kuppannagari\IEEEauthorrefmark{1}, Ji Liu\IEEEauthorrefmark{4}, Shuai Xu\IEEEauthorrefmark{1}}
\IEEEauthorblockA{\IEEEauthorrefmark{1} Department of Computer and Data Science \\ Case Western Reserve University, Cleveland, OH 44106\\
Email: \{txc461, ehh50, xxl1337, vxk285, vxc204, sxk1942, sxx214\}@case.edu}
\IEEEauthorblockA{\IEEEauthorrefmark{2} Department of Computer Science\\ Kent State University, Kent, OH 44240\\ Email: qguan@kent.edu}
\IEEEauthorblockA{\IEEEauthorrefmark{3} College of William \& Mary, Williamsburg, VA 23185 \\ Email: bren@wm.edu}
\IEEEauthorblockA{\IEEEauthorrefmark{4} Argonne National Laboratory, IL 60439 \\ Email: ji.liu@anl.gov}
\thanks{This research was supported in part by NSF Award 2216923.}}

\maketitle

\begin{abstract} 
% While much attention has been paid to the ``noisy" aspect of NISQ devices, the ``intermediate" aspect is also limiting what algorithms can be executed on them. Circuit cutting enables running circuits much larger than the underlying quantum hardware by splitting the circuit into subcircuits, running them separately, and recombining the results to receive the same bitstring distribution in the end. Previous results also show that better fidelity can be achieved using circuit cutting as compared to running full circuits on NISQ hardware. However, this process incurs overhead that scales exponentially in the number of cuts performed on the circuit. Our work introduces the idea of a ``golden cutting point" --- a location in a quantum circuit with some known property --- which allows for more efficient reconstruction of the overall bitstring distribution. Our results demonstrate speedup without any loss of accuracy in terms of the final output of the algorithm.

Quantum circuit cutting has been proposed to help execute large quantum circuits using only small and noisy machines. Intuitively, cutting a qubit wire can be thought of as classically passing information of a quantum state along each element in a basis set. As the number of cuts increase, the number of quantum degrees of freedom needed to be passed through scales exponentially. We propose a simple reduction scheme to lower the classical and quantum resources required to perform a cut. Particularly, we recognize that for some cuts, certain basis element might pass ``no information'' through the qubit wire and can effectively be neglected. We empirically demonstrate our method on circuit simulators as well as IBM quantum hardware, and we observed up to 33 percent reduction in wall time without loss of accuracy. 
\end{abstract}

\section{Introduction}
\label{sec:intro}

% \blfootnote{This research was supported in part by NSF Award 2216923.}
While quantum computers can provide exponential speed-up for certain algorithms, unveiling potential diverse and disruptive applications \cite{Grover_1996,Shor_1997,Robert_2021}, current quantum hardware lacks the scale and reliability to execute these quantum circuits. 
Preskill described such hardware as Noisy Intermediate-Scale Quantum (NISQ) computers, and much effort has been focused on increasing qubit fidelity and qubit count \cite{LaRose_2022, Abobeih_2022, Abdullah_2023, gambetta_2022}. 
Meanwhile, algorithmic developments such as variational quantum algorithms \cite{Cerezo_2021, farhi2014quantum, fedorov2022vqe} have caught substantial attention for having the potential to realize quantum advantage utilizing only NISQ machines. 
However, there are few results demonstrating provable advantage for using variational quantum algorithms over known classical algorithms~\cite{barak2021classical}.

A natural question to ask when given limited qubits is: ``Are there ways of simulating large quantum circuits using small quantum devices?''
Inspired by fragmentation methods in molecular simulations \cite{warshel1976theoretical, gordon2012fragmentation, li2007generalized, li2008fragmentation}, Peng \textit{et al.}~has demonstrated that it is possible to divide quantum circuit into subcircuits, or \textit{fragments}, that can be independently executed on small quantum computers and then recombined \cite{Peng_2020}. 
This method does not impose assumptions on circuits but requires a large amount of classical computing resources---exponential in the number of cuts---in order to reconstruct the expected output of the respective uncut circuit.

Nonetheless, quantum circuit cutting still holds great promise as an effective tool for realizing quantum algorithms.
Most notably, it has been empirically shown that reducing the circuit size, even when sufficiently large machines are available, can improve fidelity \cite{ayral2020quantum, ayral2021quantum, Perlin_2020}. 
Moreover, tailoring circuit cutting to specific applications, \textit{e.g.} combinatorial optimization \cite{saleem2021quantum} and error mitigation \cite{liu2022classical}, can effectively avoid the exponential classical overhead. 
Alternatively, stochastic methods through randomized measurement \cite{Lowe_2022} and sampling \cite{Chen_2022} have also been developed to reduce the postprocessing cost. 
In recent, state-of-the-art research, it has been empirically shown that optimization can be employed to replace traditional tomographic techniques and significantly reduce the classical resources needed for reconstruction, though at the cost of additional circuit evaluations \cite{Uchehara2022}.
Circuit cutting algorithms have also undergone improvement and variations in terms of performance.
Maximum likelihood methods have been introduced to circuit cutting to formally account for finite-shot error \cite{Perlin_2020}. 
Shadow tomography---an efficient classical representation of quantum states for predicting expectations of observables \cite{huang2020predicting}---can also be applied to individual circuit fragments \cite{chen2022quantum}. 
% There has also been work in using optimization to tune parameterized gates in a circuit so as to induce a state where some measurements can be disregarded\cite{Uchehara2022}. However, the optimization process is costly and results in more circuit executions overall.

\begin{figure*}
    \centering
    \includegraphics[width=\linewidth]{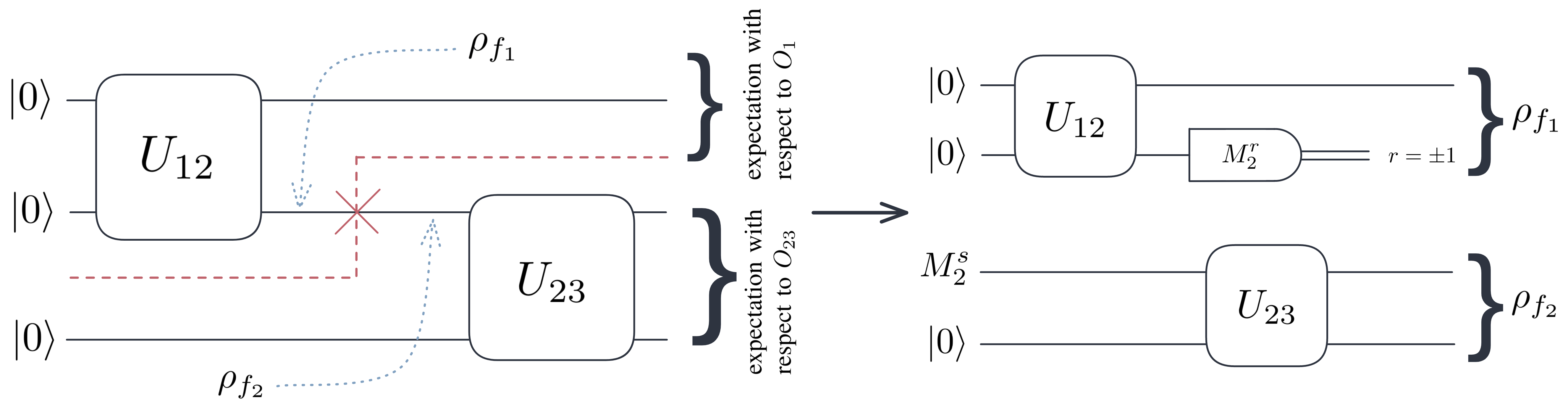}
    \caption{Circuit diagram corresponding to the 3-qubit example}
    \label{fig:circuit_diagram}
\end{figure*}

% 4. our work
Our work contributes in the direction of reducing the computational resources needed. 
In this paper, we introduce the idea of a \textit{golden cutting point}: a cut location on a quantum circuit where a certain measurement basis can be neglected, \textit{e.g.},~when the quantum state operates in a lower-dimensional subspace prior to the cut. 
Such reduction lowers the runtime of reconstructing measurement statistics from fragments by approximately 33\% for a single cut and avoids excessive circuit executions. We experimentally verify our method on both the Qiskit Aer simulator \cite{Qiskit} and superconducting quantum devices from IBM \cite{IBMQ}.

% 5. outline of the paper
The outline of this paper is as follows. 
Section \ref{sec:qccutting} presents the mathematical formulation behind quantum circuit cutting and the idea of golden cutting points. 
We begin with an elaborate three-qubit example in Section \ref{sec:example} and generalize it to arbitrary circuit bipartitions in Section \ref{sec:generalization}.
Then, we proceed to conduct numerical experiments in Section \ref{sec:experiments}. Finally, we provide insight into future directions in Section \ref{sec:conclusion}.

\section{Quantum Circuit Cutting}
\label{sec:qccutting}
% This section should be on the mathematics/theory behind previous quantum circuit cutting methods that we're building off of. I do no have the expertise to do any sort of justice to this section

Quantum circuit cutting seeks a qubit wire (or multiple qubit wires) such that, upon removing, the circuit decomposes into several independent fragments.
This is accomplished by performing quantum tomography---learning a quantum state---on the qubit upstream of the cut and recreating the inferred state in the downstream qubit. 
One can think of tomography as estimating the coefficient after expanding a quantum state with respect to some basis set, which we take to be the Pauli basis
\begin{align}
    \bigg \{ I  = \begin{pmatrix} 1 & 0 \\ 0  & 1 \end{pmatrix}, ~~&X= \begin{pmatrix} 0 & 1 \\ 1  & 0 \end{pmatrix}, \nonumber \\
    &Y  = \begin{pmatrix} 0 & -i \\ i  & 0 \end{pmatrix}, ~~Z  = \begin{pmatrix} 1 & 0 \\ 0  & -1 \end{pmatrix} \bigg \}
\end{align}
for concreteness. 
By measuring the quantum state with respect to each element in the basis, one could piece together the state prior to the cut. 
Then, by initializing the downstream qubit incident to the cut into the eigenstate of each element in the basis, one could learn the behavior of the circuit given the initial state.
Lastly, reweighting the results obtained through multiple initializations yields the behavior of an uncut circuit. 

If a circuit undergoes $K$ cuts, tomography has to be performed on all $K$ qubits at the location of the cuts. 
The set of length-$K$ Pauli strings is then an appropriate basis for a $K$-qubit system and the measurement-preparation scheme described above must be done for all elements in the set.
Thus, the cost of performing circuit cutting scales exponentially with the number of cuts.
In this section, we introduce a special class of circuits that contains a \textit{golden cutting point} and that enjoys a reduction in measurement complexity and classical postprocessing cost.
We will demonstrate this via a three-qubit example (Section \ref{sec:example}) and proceed to formalize the phenomenon for arbitrary circuit bipartitions (Section \ref{sec:generalization}).

\subsection{A Three-qubit Example} \label{sec:example}

Consider the following three-qubit state
\begin{align}
    \rho = U_{23}U_{12}|000\rangle \langle 000| U_{12}^\dagger U_{23}^\dagger
\end{align}
where unitaries $U_{12}$ and $U_{23}$ can be arbitrary quantum gates (Figure \ref{fig:circuit_diagram}). Notice that the qubit wire between the two gates can be \textit{cut}, in the sense that we can rewrite the above state as 
\begin{align}
    \rho = \frac{1}{2} \sum_{M_2 \in \mathcal B} \rho_{f_1}(M_2) \tensor \rho_{f_2}(M_2) \label{eq:cut-example-not-state}
\end{align}
where the set $\mathcal B = \{I, X, Y, Z\}$ is the Pauli basis, and
\begin{align}
    \rho_{f_1}(M_2) &= \tr_2(M_2 U_{12}\ket{00}\bra{00}U_{12}^\dagger), \\
    \rho_{f_2}(M_2) &= U_{23}(M_2 \tensor \ket 0 \bra 0) U_{23}^\dagger.
\end{align}
Here, $M_2$ refers to applying the operator $M$ onto to the second qubit. This convention will be held throughout the manuscript where numerical subscripts denote the qubit(s) an operator is acting on.

However, as currently presented, the fragments are not quantum states as Pauli matrices are traceless operators. To obtain a physical interpretation, consider the eigendecomposition eigendecomposition $M = rM^r + sM^s$ for each $M \in \mathcal B$ where $r,s \in \{+1,-1\}$ are the eigenvalues and $M^r, M^s$ are the eigenstates. Then, we can rewrite Equation \ref{eq:cut-example-not-state} as
\begin{align}
    \rho = \frac{1}{2} \sum_{\substack{M_2 \in \mathcal B \\ r,s = \pm 1}} rs \rho_{f_1}(M_2^r) \tensor \rho_{f_2}(M_2^s). \label{eq:cut-example-state}
\end{align}
The eigenstates of $M \in \mathcal B$ have unit trace and can be regarded as quantum states.

Intuitively, one can understand circuit cutting as classically passing information through the qubit wire that is cut. Inside the summation (fix an $M_2$ of choice), the information stored in the second qubit in the direction $M_2$ is measured---taking the partial trace in $\rho_{f_1}$---and passed onto the remaining circuit---initializing into state $M_2^s$. Since the \textit{circuit fragments} $\rho_{f_1}$ and $\rho_{f_2}$ can be simulated independently, through repeating the measurement-initialization scheme, we can effectively cut the original circuit, run fragments in parallel, and combine the results classically.

We are often interested in finding the expectation of the quantum state with respect to a quantum observable, $O$. Without loss of generality, assume an observable can be decomposed as $O = O_1 \tensor O_{23}$. Then, we can rewrite the expectation in terms of $\rho_{f_1}$ and $\rho_{f_2}$:
\begin{align}
    \tr(O\rho) &= \frac{1}{2} \sum_{M_2, r,s} rs ~\tr(O_1 \rho_{f_1}(M_2^r)) \tr(O_{23} \rho_{f_2}(M_2^s)) \label{eq:cut-example-expval} \\
    &= \frac{1}{2} \sum_{M_2,s} s~\tr(O_{23} \rho_{f_2}(M_2^s)) \sum_{r} r~\tr(O_1 \rho_{f_1}(M_2^r)).
\end{align}

The above summation involves 16 terms. Experimentally, the summation requires measuring the second qubit in the fragment upstream for each non-identity Pauli basis and initializing the first qubit in the downstream fragment to the respective eigenstates.

However, the computation above can potentially be reduced. Suppose that there exists an $M_{2,*} \in \mathcal B$ such that the last summation evaluates to zero, that is, 
\begin{align}
    \sum_{r=\pm 1} r~\tr\left( O_1 \rho_{f_1}(M_{2,*}^r) \right) = 0. \label{eq:cut-point-example}
\end{align}
Note that this can happen one of two ways: 
\begin{enumerate}[label=(\roman*)]
    \item Operator $O_1$ is orthogonal to the conditional state $\rho_{f_1}(M_{2,*}^r)$, \textit{i.e.}, $\tr(O \rho_{f_1}(M_{2,*}^r)) = 0$ for both $r = \pm 1$. For example, $O_1 = X$ and $U_{12}\ket{00} = (\ket{00} + \ket{11})/\sqrt{2}$, the Bell state. 
    \item The conditional state is in a subspace that ``conveys no information'' about the observable, specifically, $\tr(O_1\rho_{f_1}(M_{2,*}^r)) \neq 0$ but the weighted sum in Equation \ref{eq:cut-point-example} leads to systematic cancellations. For example, $O_1 = \ket + \bra +$ is a projector to the plus-state, and $U_{12}\ket{00}$ is again the Bell state.
\end{enumerate}
Upon noting the reduction, the number of terms in the summation drops from 16 to 12. Moreover, since any term involving $M_{2,*}$ is neglected, there is no need to estimate the expectation in the downstream fragment ($\tr \left( O_{23}\rho_{f_2}(M_{2,*}^s) \right)$), which saves circuit evaluations that use the initial state $M_{2,*}^s$. We say that the cut is a \textit{golden cutting point} if such reduction occurs.

\subsection{Generalization} \label{sec:generalization}

We formally extend the above case to more general bipartitions, but refrain from considering cutting schemes that result in more than three fragments (c.f.~\cite{chen2022quantum} for general expression of expectations with respect to circuit fragments). Suppose a $N$-qubit quantum circuit induces a state $\rho$. A set of $K$ cuts, with injective function $c$ mapping the $i$-th cut onto qubit $c(i) \in [N] = \{1,2,\dots,N\}$. By cutting the $K$ wires, the circuit can be bipartitioned into fragments $f_1$ and $f_2$. Now, cutting requires passing information from $K$ sets of basis $\mathcal B$ through the cutting point, namely, measuring upstream qubits and preparing downstream qubits with respect to operators 
\begin{align}
    \bm M = \begin{pmatrix} M_{c(1)} & M_{c(2)} & \dots & M_{c(K)} \end{pmatrix} \in \mathcal B^K.
\end{align} 
Furthermore, each operator $M$ admits spectral decomposition. Letting 
\begin{align}
    \bm r = \begin{pmatrix} r_{c(1)} & r_{c(2)} & \dots & r_{c(K)} \end{pmatrix} \in \{-1, +1\}^K
\end{align}
be a tuple of eigenvalues, we define 
\begin{align}
    \bm M^{\bm r} = \begin{pmatrix} M_{c(1)}^{r(1)} & M_{c(2)}^{r(2)} & \dots & M_{c(K)}^{r(K)} \end{pmatrix}
\end{align}
to be the $\bm r$-th eigenstate of operator $\bm M$.

Using the above notation, we can compactly write the uncut state using the fragment-induced states $\rho_{f_i}(\bm M)$ for $i = 1, 2$. Following Equation \ref{eq:cut-example-state} gives the reconstruction formula in this bipartition case: 
\begin{align}
    \rho = \frac{1}{2^K} \sum_{\substack{\bm M \in \mathcal B^K, \\ \bm r, \bm s \in \{-1,+1\}^K}} \prod_{i \in [K]} r_i s_i~ \rho_{f_1}(\bm M^{\bm r}) \tensor \rho_{f_2}(\bm M ^{\bm s}).
\end{align}
Alternatively, for any desired quantum observable $O$, suppose the operator can be decomposed to accommodate the two fragments, \textit{i.e.}, $O = O_{f_1} \tensor O_{f_2}$ up to appropriate permutation of qubit indices. This is without loss of generality as expansions using Pauli strings would yield a linear combination of operators that are qubit-wise separable. Then, we can arrive at an expression analogous to Equation \ref{eq:cut-example-expval} in terms of the fragments $\rho_{f_i}$ and their respective observables $O_{f_i}$: 
\begin{align}
    & \tr(O\rho) = \nonumber \\
    & \frac{1}{2^K} \sum_{\bm M, \bm r, \bm s} \prod_{i \in [K]} r_i s_i~ \tr \left( O_{f_1} \rho_{f_1}(\bm M^{\bm r}) \right) \tr \left( O_{f_2} \rho_{f_2}(\bm M ^{\bm s}) \right).
\end{align}
We now formally define the \textit{golden circuit cutting point}.

\begin{definition}
A $N$-qubit circuit amenable to bipartition with $K$ cuts has a \emph{golden cutting point} if there exists $k^* \in [K]$ such that 
\begin{align}
    \sum_{r_{c(k^*)}} r_{c(k^*)} ~\tr \left( O_{f_1} \rho_{f_1}(\bm M^{\bm r})\right) = 0. \label{def:golden-cut}
\end{align}
\end{definition}

Note that golden cutting points are not necessarily unique. There can be multiple cuts with negligible bases and/or multiple negligible bases in one cut. Suppose there are $K_g$ many golden cutting points and $K_r = K - K_g$ regular cutting points, the run time complexity of reconstructing the expectation in this bipartite case scales as $\mathcal O(4^{K_r}3^{K_g})$. Moreover, there is no need of preparing the downstream fragments into the respective eigenstates, reducing the number of circuit evaluations from $\mathcal O(6^K)$ to $\mathcal O(6^{K_r}4^{K_g})$. It is worth noting that the eigenstate preparation scheme is overcomplete and alternative bases, \textit{e.g.} the symmetric informationally-complete (SICC) basis, can be used to achieve $\mathcal O(4^K)$ circuit evaluations without invoking golden circuit cutting formalism. However, employing the SICC basis would require more involved implementation, namely, solving linear systems, in order to construct the appropriate tensor during reconstruction. 

We emphasize that golden cutting points as its current form is strictly a product of circuit design. That is, we can design circuits that admits golden cutting points at known locations (c.f.~Section \ref{sec:experiments}). In reality, the existence of golden cutting points depends on the choice of the observable $O$, how it can be decomposed into form amendable to cutting, and the the state of the qubit prior to the cut. It is unlikely that an arbitrary circuit would exhibit such property without intricate designs, and we defer further discussion on finding golden cutting points to Section \ref{sec:conclusion}.

% Golden cutting point extend to arbitrary, multipartite cutting schemes. \red{(going into tensor contractions, but it might be confusing)}.

\section{Experiments}
\label{sec:experiments}

\begin{figure}
    \centering
    \includegraphics[width=0.5\textwidth]{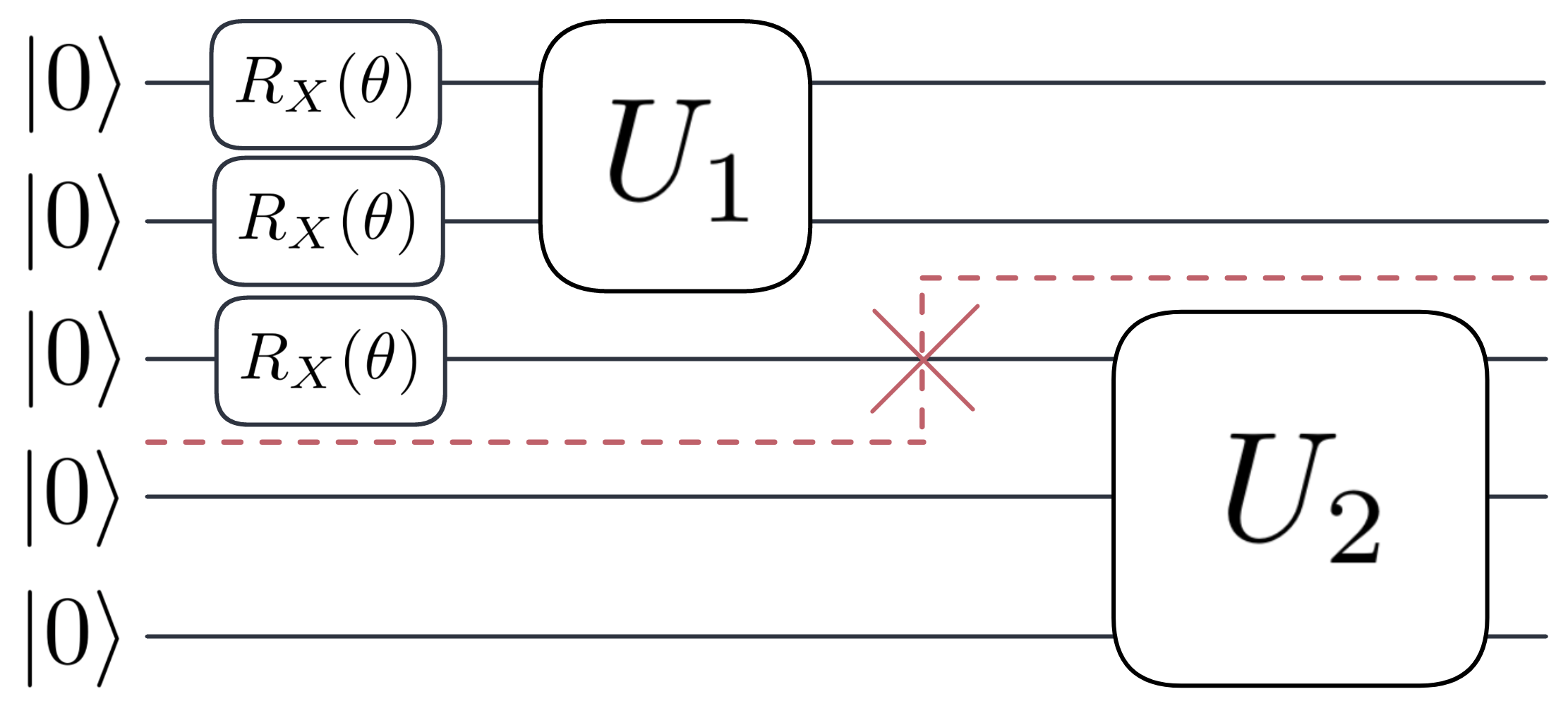}
    \caption{Example 5-qubit circuit with golden cutting point. $U_1$ and $U_2$ are randomized circuits. Note that other odd-number circuit widths were used too, mainly 7-qubits split into subcircuits of 4 qubits each.}
    \label{fig:5_qubit_experiment}
\end{figure}

We demonstrate the capability of golden cutting points numerically. Specifically, experiments were designed to verify two aspects of our work: first, that our method does not sacrifice the correctness of reconstruction and second, that our method reduces the overall runtime of the algorithm. 
The circuit used in our experiment takes the form showed in Figure \ref{fig:5_qubit_experiment} where the sizes of the circuits were tailored to the size of the device it was run on. Most commonly, we split 5- and 7-qubit circuits into fragments of 3 and 4 qubits respectively. In addition to quantum hardware, we also tested our method on the Aer simulator from Qiskit \cite{Qiskit}. 

The generated circuits featured collections of $RX$ gates with the rotation angle $\theta$ was chosen uniformly at random from the interval $[0, 6.28]$, as well as random gates generated using the \verb|random_circuit()| function in Qiskit, denoted by $U_1$ and $U_2$ in Figure \ref{fig:5_qubit_experiment}. In our experiments, we want to acquire the bitstring probability distribution generated by repeated measurements in the computational ($Z$) basis. Alternatively, to match the formalism of Section \ref{sec:generalization}, for all bitstrings $\hat b \in \{0,1\}^n$, one can interpret the above as estimating the expectation of the projector observable $\Pi^{\hat b} = \ket{\hat b} \bra{\hat b}$ where $n=5, 7$ is the number of qubits in a circuit. This observable decomposes straightforwardly upon cutting:
\begin{align}
    \Pi^{\hat b} = \Pi^{\hat b_1} \tensor \Pi^{\hat b_2}
\end{align}
where $\hat b_1$ and $\hat b_2$ are bitstrings of length $\lfloor n/2 \rfloor$ and $\lceil n/2 \rceil$ respectively such that the concatenation recovers $\hat b$. By repeating many measurements in the computational basis and obtaining a bitstring distribution, we can then estimate the expectation of the projector observables.
% Following the notation shown in Figure \ref{fig:circuit_diagram} and elaborated on in Section \ref{sec:example}, our observable $O$ for these experiments was $Z^{\otimes n}$ (measure in the standard basis) and $O_1$ and $O_2$ were $Z^{\otimes \lfloor n/2 \rfloor}$ and $Z^{\otimes \lceil n/2 \rceil}$ respectively.
The restrictions imposed on our circuit ansatz creates a golden cutting point. In particular, the contribution of first fragment to the total expectation (with respect to the projector operator above, which has only diagonal components) conditioned on observing each eigenstate of the Pauli $Y$ operator leads to components of equal magnitudes. Once weighed by the respective eigenvalues, the trace terms systematically cancels and induces a golden cutting point as defined in \ref{def:golden-cut}.
% Note that this satisfies the first way \ref{def:golden-cut} can evaluate to 0. That is, this state is orthogonal to $Y$, so measuring in that basis gives a golden cutting point.

\subsection{Verifying Reconstruction Accuracy}

\begin{figure}
    \centering
    \includegraphics[width=0.5\textwidth]{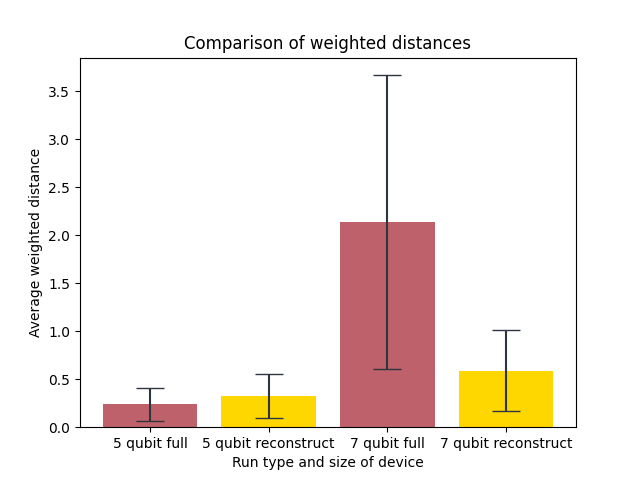}
    \caption{Comparison of weighted distances for uncut circuit evaluations and reconstructed meausurements from fragment data. Each bar was averaged over 10 independent trials. Each trial consisted of 10,000 shots per fragment. Error bars represent 95\% confidence intervals.}
    \label{fig:distances}
\end{figure}

We first proceed to verify the correctness of our golden cutting point method.
We performed noiseless (in the sense of hardware noise) simulations of the full uncut circuit to obtain the ground truth distribution and compared our golden cutting point method to it.
In addition, we ran the same circuits, both uncut and fragmented, on superconducting IBM Quantum devices \cite{IBMQ}.
To compare distributions, we now introduce a \textit{weighted distance function} $d_w$:
\begin{align}
    d_w(p; q) = \sum_{x \in \mathcal X} \frac{\left( p(x) - q(x) \right)^2}{q(x)}.
\end{align}
for distributions $p$ and $q$ with support $\mathcal X$.  
In the experiment, $p$ was our ``test'' distribution and $q$ was the ``ground truth.'' 
This weighted distance function penalizes large percentage deviations more than other metrics such as the total variational distance.

Using this weighted distance function, we determined the distance between the ground truth bitstring distribution from the Aer simulator and the distribution obtained by running the full circuit on quantum hardware. We then also determined the distance between the ground truth distribution and the distribution obtained by running our golden cutting method on quantum hardware. We repeated the experiment for machines of two different sizes---a 5-qubit device would run a 5-qubit circuit that is split into two 3-qubit subcircuits and a 7-qubit device would run a 7-qubit circuit split into two 4-qubit subcircuits. 
Each circuit configuration was repeated for 10 trials with each trial consisting of 10,000 shots per (sub)circuit.
The results of the experiments are shown in Figure \ref{fig:distances}.

Because previous work has found that quantum circuit cutting has yielded a benefit in terms of fidelity, we expected similar results. We were surprised to find that such a benefit is non-existent on 5-qubit devices and not detectable within 95\% confidence intervals on a 7-qubit device. This is likely due to the fact that our circuits were not particularly deep---only a few gates in each. As the circuit depth increases we should expect to see more of a benefit from circuit cutting. Additionally, an average of only 10 trials was performed, leaving our uncertainties relatively large. In our future work we will run more trials, potentially on larger devices, or with greater depths, to more precisely determine the true distance for this method. Regardless, in this work, we have verified that (within error) our method performs as well as full circuit execution on real hardware in terms of outputting the correct bitstring distribution.

\begin{figure}
\centering
\includegraphics[width=0.5\textwidth]{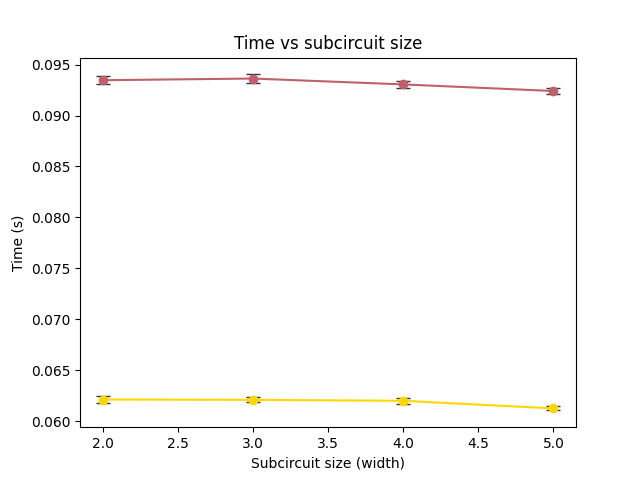}
\caption{Runtime comparison of circuits with (in gold) and without (in red) golden circuit cutting optimization. Each configuration was repeated for 1000 trials and each trial involved 1000 shots for each (sub)circuit. Error bars represent 95\% confidence intervals.}
\label{fig:sim_runtime}
\end{figure}

\subsection{Algorithm runtime}

The presented golden cutting point method predicts a reduction in the number of measurements, thereby lowering the runtime of the procedure. 
In this experiment, we recorded the time taken for gathering fragment data and reconstructing them on a randomly generated circuit. 
We assumed the golden cutting point was known a priori (see Section \ref{sec:conclusion} for determining the existence of golden cutting points), and studied the runtime reduction from exploiting this knowledge.
The results are illustrated in Figure \ref{fig:sim_runtime}.

\begin{figure}
    \centering
    \includegraphics[width=0.5\textwidth]{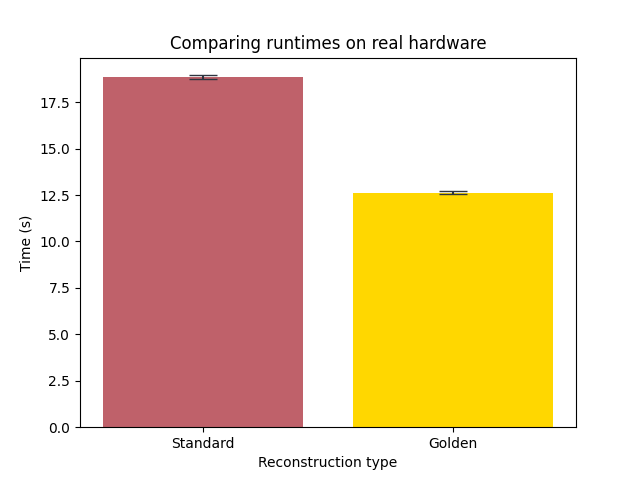}
    \caption{Circuit cutting runtime with and without golden cutting point on quantum devices from IBM. Each bar represents the average runtime of 50 trials and each trial comprised of 1,000 shots for each (sub)circuit.}
    \label{fig:hardware_runtime}
\end{figure}

We then repeated the same experiment using real devices available through the IBM Quantum Experience \cite{IBMQ}. 
We report a speedup of 33 percent using our method as compared to the standard method~\cite{ayral2021quantum}, as shown in Figure \ref{fig:hardware_runtime}. Specifically, we find that the average time for execution using the standard (without golden cutting point consideration) reconstruction method was 18.84 seconds whereas the golden cutting point method had a mean of 12.61 seconds. This reduction in run time is largely attributed to the reduced number of circuit evaluations. Particularly, we avoided having to execute a third of the total shots by neglecting one basis element, bringing the total number of circuit executions down from $4.5 \times 10^{5}$ to $3.0 \times 10^{5}$. This demonstrates the applicability of our method: it is simple to implement and produces a considerable reduction in wall time needed to perform circuit reconstruction. 
% However, the majority of the speedup at this stage (small $K$) comes from being able to neglect certain bases to measure and prepare in, rather than avoiding the reconstruction overhead. We ran a total of 450,000 shots to time the standard method (50 trials $\times$ 9 circuits/trial $\times$ 1000 shots/circuit), but only 300,000 shots to time the golden cutting method (50 trials $\times$ 6 circuits/trial $\times$ 1000 shots/circuit). This is important because this relatively simple method can produce a considerable reduction in the wall time needed to perform circuit reconstruction.

\section{Conclusion}
\label{sec:conclusion}

In this paper, we introduced the idea of a golden cutting point where elements of a basis set can be neglected (Section \ref{sec:generalization}). Neglecting these basis elements allows for a considerable decrease in the runtime. 
The reduction can be attributed to (1) fewer terms are involved when combining measurement results from fragments, and (2) fewer circuit evaluations are needed in the fragment downstream of the golden cut. We verified the correctness and demonstrated the lowered runtime in both simulators and real quantum hardware (Section \ref{sec:experiments}).
We observed that golden cutting points do not come at the cost of accuracy, and a statistically significant 33\% decrease in runtime persists in both a simulated environment and quantum devices.

The introduction of the golden cutting point begs the question of applicable algorithms that have built-in golden cutting points. Variational circuits are a potential candidate due to their flexibility in the circuit ansatz. Although variational algorithms for combinatorial optimization or chemical simulations have great structural assumptions given by the Hamiltonian, quantum machine learning circuits typically do not possess these constraints and therefore are likely more amenable to exploiting the golden cutting point formalism. 

Moreover, detection of the existence of a golden cutting point is also an interesting one. In this work, we assumed the golden cutting point was known a priori. However, it is not obvious whether a basis can be neglected without simulating the circuit. Thus, we suspect there are methods for detecting golden cutting points ``online'' during the execution of the circuit cutting procedure through sequential empirical measurements. Note that this does not necessarily sacrifice the parallelism of circuit cutting as golden cutting points only affect the fragments directly incident to the cut. However, performing such a method would require further statistical analysis of acceptable error and the amplification of error through tensor contraction. 

\section*{Acknowledgements}

We acknowledge the use of IBM Quantum services for this work. The views expressed are those of the authors, and do not reflect the official policy or position of IBM or the IBM Quantum team. This material is based upon work supported by the U.S. Department of Energy, Office of Science, National Quantum Information Science Research Centers. The colors used in figures were based on Nord Theme \cite{nord}.
Daniel T. Chen and Ethan H. Hansen contributed equally to this work.

\bibliographystyle{IEEEtran}
\bibliography{IEEEabrv,ref}

\end{document}